\documentclass[10pt,a4paper]{article}
\usepackage{graphicx}

\usepackage{authblk}
\usepackage{fullpage,amsfonts,setspace}

\usepackage{subcaption}

\usepackage{amstext}
\usepackage{url}

\title{Measuring Accessibility using Gravity and Radiation Models }

\author[1,2,*]{Duccio Piovani}
\author[1]{ Elsa Arcaute }
\author[1,4]{Gabriela Uchoa }
\author[1,3]{Alan Wilson}
\author[1]{Michael Batty}

\affil[1]{\small{Centre for Advanced Spatial Analysis (CASA), University College London (UCL), 90 Tottenham Court Road , London, W1T 4TJ}}
\affil[2]{\small{nam.r , 3 rue de Steinkerque, 75018 Paris}}
\affil[3]{\small{The Alan Turing Institute, British Library, 96 Euston Road, London NW1}}
\affil[4]{\small{Prefeitura Municipal de Teresina, Pra\c{c}a Marechal Deodoro da Fonseca 860,  Teresina, Brasil  64000-160\\}}

\affil[*]{duccio.piovani@gmail.com}

\begin{document}

\maketitle
\begin{abstract}
Since the presentation of the Radiation Model, much work has been done to compare its findings with those obtained from Gravitational Models. These comparisons always aim at measuring the accuracy with which the models reproduce the mobility described by origin-destination matrices. This has been done at different spatial scales using different datasets, and several versions of the models have been proposed to adjust to various spatial systems. However the models, to our knowledge, have never been compared with respect to policy testing scenarios. For this reason, here we use the models to analyze the impact of the introduction of a new transportation network, a Bus Rapid Transport system, in the city of Teresina in Brazil. We do this by measuring the estimated variation in the trip distribution, and formulate an accessibility to employment indicator for the different zones of the city. By comparing the results obtained with the two approaches, we are able, not only to better assess the goodness of fit and the impact of this intervention, but also to understand reasons for the systematic similarities and differences in their predictions. 
\end{abstract}

\section*{Introduction}

Assessing the impact of new infrastructural projects is a challenging and demanding task that requires knowledge or estimates of the mobility of the individuals living in the city.  Many models have been developed to this effect\cite{stouffer1940intervening,anas1983discrete,ben1985discrete,Wilson1969},
focusing on different scales of the urban system, according to the quality of the data available. Traditionally, these models allocate trips from one geographical zone to another, according to estimates of where people live and work. Infrastructure projects are then assessed following changes in accessibility which are computed from the model's predictions of the people living and working in these zones. Among the models that have been proposed over the decades, the gravitational model \cite{Wilson1969} has been one of the most widely adopted in various contexts (for example,   \cite{jung2008gravity,anderson2011gravity,khadaroo2008role,krings2009urban,balcan2009multiscale,wilson2008boltzmann}). Depending on the  information available on the demographics and mobility of the individuals, this model exists in the following different forms as an unconstrained, singly constrained, and  doubly constrained model. Naturally, the more information available to calibrate the model, the better its performance can be assessed against observed data, notwithstanding the fact that information is not always available for this purpose.

Recent years have seen a dramatic increase in available data on individuals and their urban environments, allowing researchers to test  these models more effectively, thus providing more detail on the outstanding problems of human mobility. This has  prompted a surge in the literature, where new models have been proposed, such as the \emph{Radiation Model} \cite{Simini2012}. This model takes its inspiration from the intervening opportunities model \cite{stouffer1940intervening} where flows are modelled without parameters and take only as input the population distribution. 
It produces predictions with a high degree of accuracy at the intra-county scale, hence introducing a new benchmark in the field of mobility modelling. This has triggered the interest of many researchers, and many works have appeared where its predictions are compared with those of the more traditional gravity model \cite{kang2015generalized, Simini2013,liang2013unraveling,Lenormand2012,Lenormand2016,Yang2014,Masucci2013}.
These efforts have focused on comparing the accuracy of the models in reproducing observed origin-destination matrices. As shown in \cite{Lenormand2012, Yang2014,Masucci2013} the main limitations of the Radiation Model is in its inability to produce adequate outcomes at different spatial scales which is a direct consequence of its own virtue of being parameter free. In order to overcome this limitation, several solutions have been proposed: notably in \cite{Masucci2013} the authors introduce a \emph{normalised} version of the model in order to take into account of finite size effects, while in \cite{Yang2014} the authors propose an \emph{Extended} version of the Model introducing a parameter that can be calibrated to the data. These studies show how in general, the accuracy in reproducing the observed flows of the \emph{extended} and the \emph{normalised} version are often comparable to those obtained from the doubly-constrained gravitational model.

The origin-destination matrices used to compare the models in previous works, are extracted both by conventional datasets, as for example mobility surveys and census data, or unconventional datasets such as mobile phone or geo-located social media data. Very often though the matrices are outdated, incomplete or obtained by biased samples of the population \cite{lenormand2014cross, }. We therefore believe that the models need to be compared in a more practical manner where their results can be read and interpreted more easily. For this reason, here we want to take a different approach and exploit the two different models to quantify and predict the impact of the introduction of a new Bus Rapid Transit (BRT) system in the city of Teresina in Brazil. BRT systems are increasing in popularity worldwide as an alternative cost-effective investment in comparison to expensive urban rail transport projects \cite{cervero2013bus}. Readers can see from http://brtdata.org/ that there are more than 206 cities which have introduced some kind of BRT system and with the number of new corridors under construction increasing steadily. Indeed emerging economies have been seduced by the publicized BRT success from cities as Curitiba and Bogot\`{a} \cite{rodriguez2004value, rodriguez2008land,lindau2010curitiba,levinson2002bus}, cities that developed high performance with BRT systems, enhancing mobility and sustainability at an affordable price. For this reason,  implementations of BRT systems are now being evaluated by many cities \cite{cervero2011bus,deng2013bus} and there are many studies of these proposals\cite{hensher2008bus,currie2005demand,day2014does}. With this in mind, the main goal of this paper is to use both the Gravitational and the Radiation models to quantify the effects of such transportation interventions, measuring the variation in accessibility, and comparing the two results in terms of an analysis of their similarities and differences. 

\section{The Case Study: The BRT implementation in Teresina}

\begin{figure}[t!]
\centering
\includegraphics[scale=0.23]{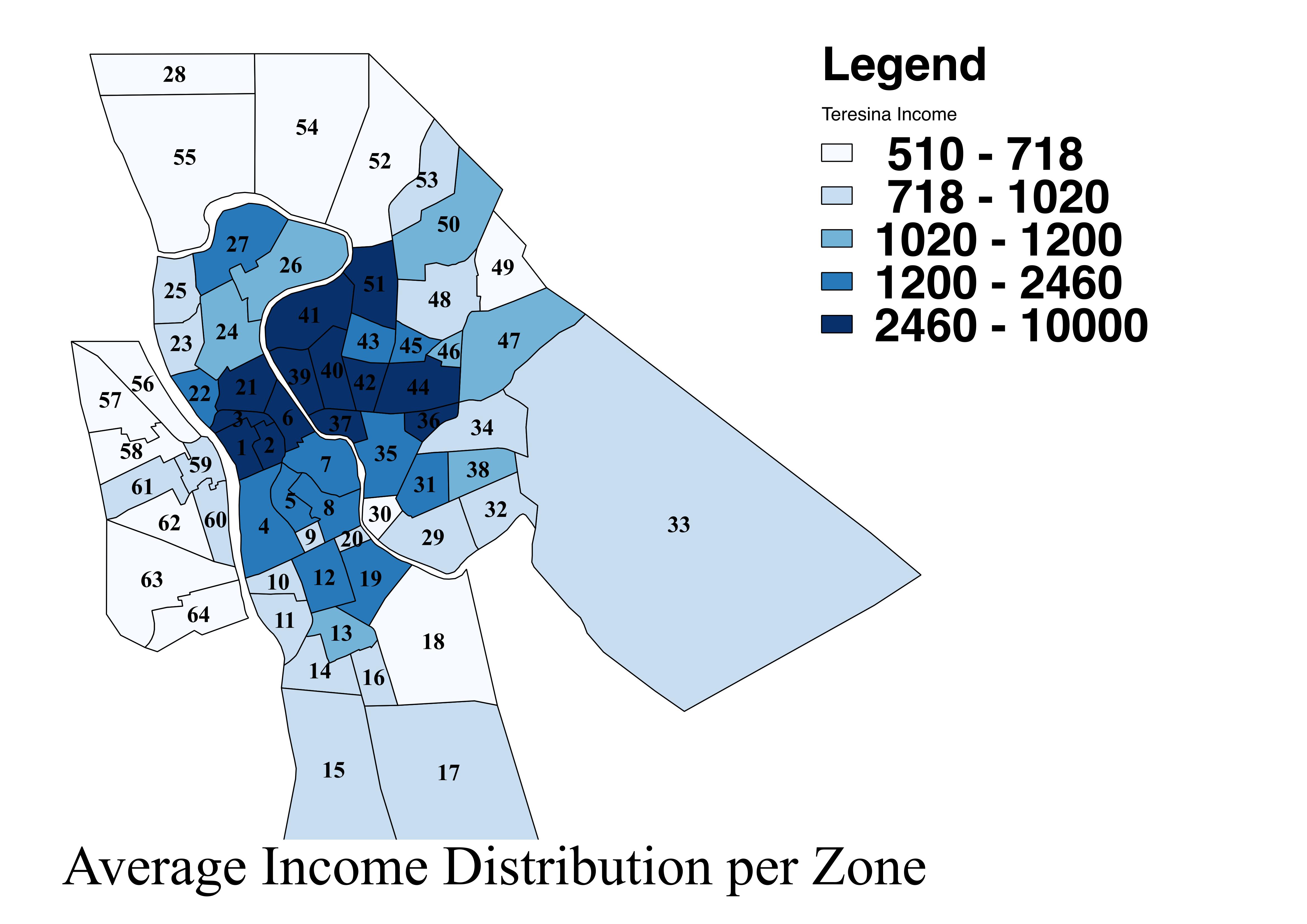}
\includegraphics[scale=0.23]{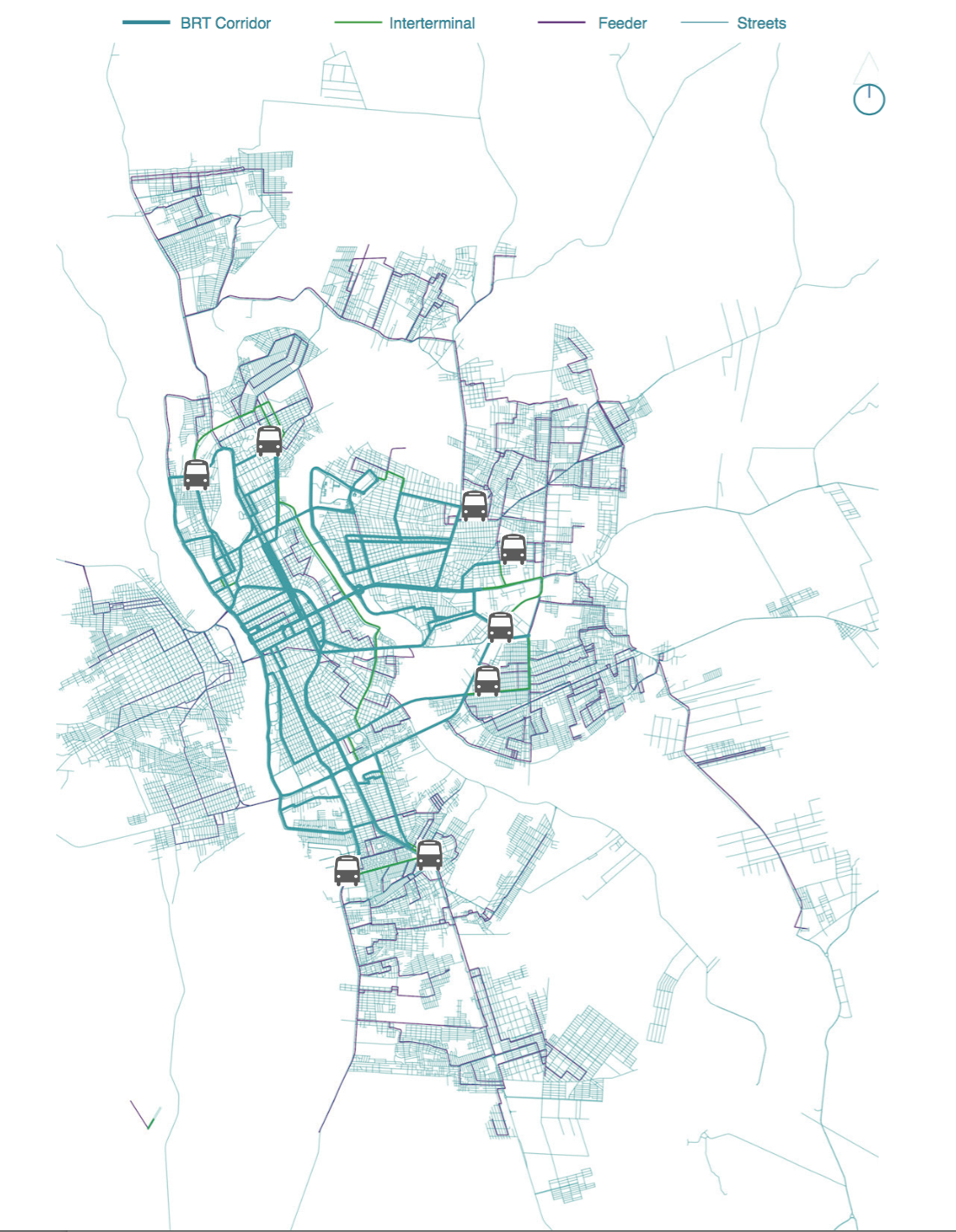}
\caption{ (Left Panel) A map of Teresina, zones 1 to 55, and Timon, zones 56 to 64, where zones are coloured according to their average income from white, low income,  to dark blue, high income. (Right panel) A map of the new BRT stops and their corridors. The BRT stops are only found in Teresina but  mobility in Timon is clearly affected by the scheme. }
\label{fig:1}
\end{figure}

As mentioned above, our case study will be the city of Teresina in  Brazil, a medium size metropolis which is currently implementing a BRT system. It is the capital of the state of Piau\`{i} and its metropolitan region has almost $1.2\text{x}10^6$ inhabitants according to the last census estimate made by the "Instituto Brasileiro de Geografia e Estat\'{i}stica"in 2015. The administrative region, named "Regi\~{a}o Integrada de Desenvolvimento da Grande Teresina - RIDE/Grand Teresina", is composed of 15 municipalities. However, only two municipalities are served by a connected urban transport network - Teresina and Timon (see Fig.~\ref{fig:1}). Both cities together concentrate most of the population in the region with just over $10^6$ inhabitants. The metropolitan public transportation consists of a bus network, and a rail service connecting the southeast zone to the city centre. The rail service operates sharing the infrastructure with freight trains on a single track in both directions, resulting in a sparse and low usage service. For this reason, it will not be considered for cost comparisons in this study. The present structure of the public transport system is thus non-hierarchical. The majority of the lines form a radial scheme, departing from the suburbs towards the city centre with a  few services that directly connect zones in the suburbs. In recent years, several exclusive bus lanes in the central area have been constructed which aim to reduce travel time on congested roads.
In 2008, the municipality of Teresina approved its Transport and Mobility Master Plan which proposes a new system of public transport in the city. It relies on an Origin-Destination survey of trip making conducted in 2007 which was developed to analyse travel patterns and predict future scenarios \cite{Strano2008}. The  original proposal suggests the implementation of a Bus Rapid Transit system (BRT), splitting the existing routes and creating a hierarchical system of feeder, inter-terminal and trunk services.  This will be composed of eleven terminals connected through express bus corridors. 
The proposal aims to increase the effectiveness of public transportation in the city, improving the accessibility to jobs, education and public services. In this paper, through different measures of accessibilities, we evaluate the impact of such an infrastructure project.

\section{Methods}

In this section we will present and briefly recount the details of the models we have used to estimate the impact of the introduction of the BRT system in Teresina and Timon, while also introducing the equations we have used to measure the accessibilities. As we will see, we have taken into account the journey to work distribution using both models and infrastructure to calculate the accessibilities. 

\subsection{The Gravity Model}

In the gravity model approach, the flow from zone $i$ to zone $j$ is proportional to the opportunities,employment in this case, $E_j$ in destination $j$, and to the demand in origin $i$ (which is represented by the population) $P_i$, and weighted by the cost function $f(c_{ij})$. Given our approach, we consider the cost $c_{ij}$ of going from $i$ to $j$ as the expected time of travel using the public transportation network (see appendix for details on the travel time calculations), and the function as being an exponential decay of the form $f(c_{ij}) = \text{e}^{-\beta c_{ij}}$, where $\beta$ is a parameter that has to be calibrated on data. Traditionally to model the journey to work trip distribution, the total number commuters is constrained (outflow), as are the employees arriving at work (inflow). This corresponds to the doubly-constrained model, where the flows are described by the equation
\begin{equation}
T^\text{dbl}_{ij}  = A_i B_j P_{i} E_{j}  \text{e}^{-\beta c_{ij}}
\end{equation}
where $A_i$ and $B_j$ are two normalization constants that one has to solve iteratively. By imposing the constraints on the total outflow and on the total inflow
\begin{equation}
\sum_j T_{ij} = P_i \label{out_constraint} \qquad \qquad \qquad \sum_i T_{ij} = E_j
\end{equation}
and following the procedure in \cite{Wilson1969}, we get  
\begin{equation}
A_i = \frac{1}{\sum_k B_k E_k \text{e}^{-\beta c_{ik}}} \qquad \qquad  B_j = \frac{1}{\sum_k A_k P_k \text{e}^{-\beta c_{kj}}}
\end{equation}
In its single constrained version, where only the constrain on the outflow is kept, the flow from origin $i$ and destination $j$ in this case has the form
\begin{equation}
T^\text{sng}_{ij} = Z_i P_i E_j \text{e}^{-\beta c_{ij}} 
\end{equation}
where $Z_i$ is the normalisation constant. Imposing the constraint on the outflow $\sum_j T_{ij} = P_i$ leads to
\begin{equation}
Z_i = \frac{1}{\sum_k E_k \text{e}^{-\beta c_{ik}}}
\label{eq:Z}
\end{equation}
In the next section we will use both forms to reproduce the origin-destination matrix and to calculate the accessibilities. 

\subsection{The Radiation Model}

The original radiation model \cite{Simini2012} takes its inspiration from the intervening opportunities model \cite{stouffer1940intervening}. In this 
approach  the probability of commuting between two units $i$ and $j$  depends on the number of opportunities between the origin and the destination, rather than on their distance. In its original formulation the radiation model made use of the population in each zone, using it also as a proxy for employment. The flow between zone $i$ and $j$ is therefore quantified as
\begin{equation}
T^\text{rad}_{ij} = T_i  \frac{P_i P_j}{(P_i + P_{ij})(P_i+P_j+P_{ij})} 
\label{eq:Rad_0}
\end{equation}
where $P_{ij}$ is the population in zones included in a radius of distance $d_{ij}$ (and excluding those of zones $P_i$ and $P_j$). These  represent the opportunities between them, and $T_i$ is the amount of commuters in $i$. As presented, the model in eq.(\ref{eq:Rad_0}) was formulated to describe flows happening on large scales and the absence of parameters to calibrate makes the model hard to fit to smaller scales. For this reason, the form we have used is slightly different and following \cite{Masucci2013} we have added a normalisation constant that takes into account the \emph{finite size} of the system. Moreover we have used the employment to characterize opportunities rather than the population. The flows between zones $i$ and $j$ are now described by 
\begin{equation}
T^\text{rad}_{ij} = \frac{P_i}{(1-\frac{P_i}{P})} \frac{E_i E_j}{(E_i + E_{ij})(E_i + E_j + E_{ij})}
\label{eq:NormRad}
\end{equation}
where $P$ is the total amount of population in the system, and where $E_{ij}$ is the amount of employment between zones $i$ and $j$.

We have also used the $\emph{Extended}$ version of this model presented in \cite{Yang2014} where a parameter $\alpha$ is introduced, and whose calibration makes the model adaptable to different spatial scales. The flow in this version is derived by combining the original Radiation model with \emph{survival analysis}\cite{miller2011survival} and in this context the probability of commuting from $i$ to $j$ is described by the equation
\begin{equation}
P(1 | E_i,E_j,E_{ij})  =  \frac{\left[(E_i + E_j +E_{ij})^\alpha - (E_i+E_{ij})^\alpha) \right](E_i ^\alpha + 1)}{\left[ (E_i+E_{ij})^\alpha+1\right] \left[(E_i+E_j +E_{ij})^\alpha +1\right]}
\label{eq:ExRad1}
\end{equation}
The details of the calculations that lead to this form may be found in \cite{Yang2014}. The flows in the extended version are the product eq.(\ref{eq:ExRad1}),  the population in the origin zone $i$, and normalisation term
\begin{equation}
T^\text{ext}_{ij} = P_i \frac{P(1 | E_i,E_j,E_{ij})}{\sum_k P(1 | E_i,E_k,E_{ik})}
\label{eq:ExRad}
\end{equation}
One may notice that as per our construction, eq.(\ref{eq:ExRad}) is constrained to meet the outflow $\sum_k T_{ik} = P_i$, but not  the inflow $\sum_k T_{ki} \neq E_i$ exactly as in the singly-constrained version of the gravity model. 

\subsection{Measures of Accessibility}

As mentioned in the introduction, in order to quantify the impact of the new infrastructure, we measure the accessibility predicted by the models before and after the introduction of the BRT, which is done by using both $c^\text{old}_{ij}$ and the updated cost matrix $c^\text{brt}_{ij}$. Accessibility has become a central concept in physical planning in the past decades \cite{bertolini2005sustainable,vandenbulcke2009mapping,geurs2004accessibility,batty2009accessibility,jones1981accessibility}, and many different definitions exist which depend on the specific application. In general as stated in \cite{batty2009accessibility}, accessibility associates some measure of opportunity at a place with the cost of actually realising that opportunity, or in other words as the cost of getting to some place traded off against the benefits received once that place is reached. In \cite{batty2009accessibility}, two main types of accessibility are defined: type 1 takes into account the locational behaviour described through models, with  infrastructure only implicitly considered;  type 2, considers the physical infrastructure and some generalised measure of the \emph{distance} from the zone of interest to all others.  For a zone $i$ we define accessibility of type 1 as 
\begin{equation}
A^{1}_i  = \frac{\sum_j T_{ij} (1/c_{ij})}{\sum_j T_{ij}} 
\label{eq:A1}
\end{equation}
where $c_{ij}$ is the cost of commuting from $i \rightarrow j$ (this is a measured quantity which depends on the infrastructure), and $T_{ij}$ is the predicted flow and depends on the model used to calculate it.  The accessibility in eq.(\ref{eq:A1}) quantifies the inverse of the average cost of commuting from the given area: high values of $A_i$ correspond to the flows happening with low values of $c_{ij}$ and viceversa. We define, as a simple measurement of the infrastructure, the accessibility of type 2 as 
\begin{equation}
A^{2}_i = \frac{1}{N_d}\sum_j \frac{E_j}{c_{ij}}  
\label{eq:A2}
\end{equation}
where $E_j$ are the opportunities (employment) in zone $j$, and $N_d$ is the total number of destinations. Once again eq.(\ref{eq:A2}) is telling us the average benefit-cost ratio for zone $i$. As we can see, there is no modelling involved in this measure, and we have only exploited the cost matrix and the opportunities distribution. Comparing  eq.(\ref{eq:A1}) and eq.(\ref{eq:A2}) allows us to understand the information benefit of adding a modelling layer to the analysis, especially given the use of the different kinds of model we have applied.

\section{Results}

In the first part of this section, we will produce the trip duration distribution with the different models and calibrate it to the trip data (details of which with respect to the origin-destination matrix data are to be found in the appendix). As said, the reproduction of the observed origin destination matrices is usually how the goodness of fit of such a model is tested. From this analysis, it will be clear that the normalized radiation model is not appropriate in fitting mobility problems at this scale (as it has not been developed with intra city trips in mind), and will therefore, we will not consider this model further. In the second part, we will compare in great detail the results obtained when calculating the impact on the accessibility of the new BRT bus network. We will do this only using the extended version of the radiation and the gravitational models. Finally, to understand these results, we study the characteristics of the zones whose predictions are similar using the two approaches and compare them with the zones for which they differ. 

\subsection{Trip Duration Distributions}

To calibrate the parameters of the models we have used we used the maximum likelihood method. As one can see in the appendix the origin-destination survey counts only 138 households, and therefore does not contain information on trips from every zone to every other. For each origin zone, we only have information on trips to a limited number of destinations. This implies that we cannot calibrate the models using  the common part of commuters based on the Sorensen's index \cite{sorensen1948method} as done in previous comparisons \cite{Lenormand2012 , Lenormand2016, Yang2014, Masucci2013}. We therefore extract the trip duration distribution of the bus trips contained in the dataset \cite{Strano2008} , $p_\text{obs}(t)$ and for each value of the parameter which we generically indicate as $p$, we measure the distribution of the predicted duration of the trips $p_\text{mod}(t,p)$. We then find the values of the parameters that maximise the likelihood expression
\begin{equation}
L(p) = \sum_t p_\text{obs}(t) \log{(p_\text{mod}(t,p))}
\label{eq:L}
\end{equation}
where $p=\alpha$ for the extended radiation and $p=\beta$ for the gravity model.
\begin{figure}
\centering
\includegraphics[width=0.8 \linewidth]{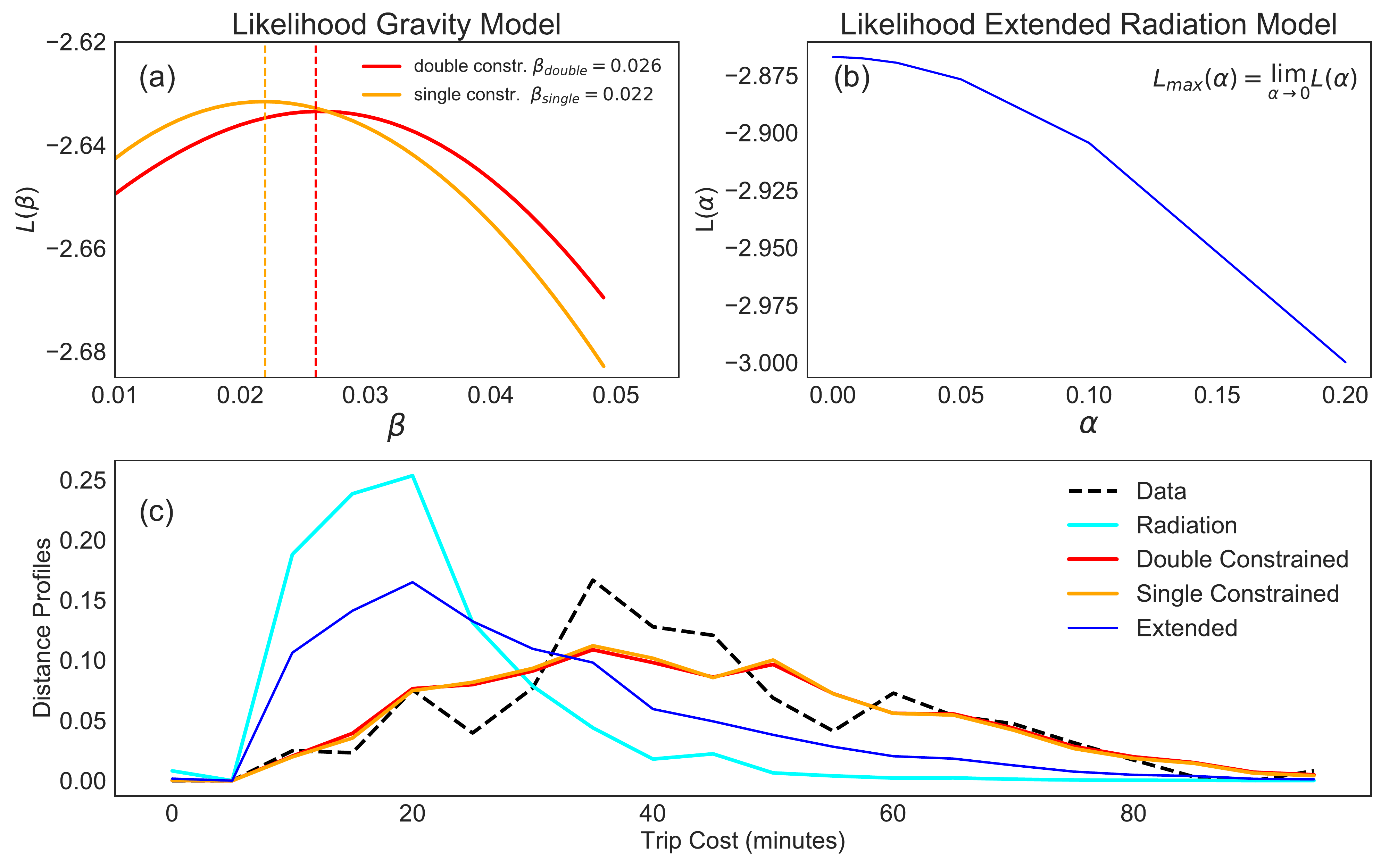}
\caption{In the top panels \textbf{(a)} and \textbf{(b)} we show the behaviour of the likelihood for the gravity and the extended radiation models. If the gravity models exhibit a clear maximum for $\beta_{sng} = 0.022$, $\beta_{dbl} = 0.026$, in the extended radiation context this is reached asymptotically for $\alpha \rightarrow 0$. In \textbf{(c)} we compare the trip duration distribution found in the data with what comes out from the models.}
\label{fig:2}
\end{figure}
In fig.(\ref{fig:2})(a)-(b) we show the form of the likelihood for the doubly and singly constrained gravity model, and for the extended radiation model. For both versions of the gravity approach the likelihood presents a clear maximum with $\beta_{single} = 0.022$ and $\beta_{double} = 0.026$, values we have used to produce the results shown in this paper. 

When repeating the calculation in eq.(\ref{eq:L}) for the extended radiation model, we see how the maximum is asymptotically reached for $\alpha \rightarrow 0$. Given that we are working at an intra-city scale this is not surprising, and in the supplementary material of \cite{Yang2014}(in section 9) the authors have solved the model's form as eq.(\ref{eq:ExRad}) for this limit with these scales in mind. The equation that describes the flow from zone i to zone j for $\alpha \rightarrow 0$ becomes 
\begin{equation}
\lim_{\alpha \rightarrow 0} T^\text{ext}_{ij} =  P_i \frac{n_j / (n_i + s_{ij})}{\sum_k n_k / (n_i + s_{ik})}
\label{eq:ext}
\end{equation}
In fig.(\ref{fig:2})(c) we show the trip duration distribution obtained with the different models. The black dashed curve indicates the trip duration distribution found in the data $p_\text{obs}(t)$, the red and orange curves represent the singly and doubly-constrained  gravity models, the normalised and extended radiation the cyan and blue curves.

It clear that with no calibration process, the normalised radiation is not in agreement with what is seen in the data, and how  the distribution obtained with the extended version of the model, in the limit $\alpha \rightarrow 0$, moves closer to what seen in the data. If on the one hand, this indicates that indeed the extended version of the radiation model makes it a better fit for simulating mobility at small scales, these results seem to be yet another proof of superiority of the gravity approach when dealing with urban mobility problems. We will therefore exclude the normalised radiation from the following analysis, and concentrate on the radiation model in its extended form. 

\subsection{Accessibility Variations after the Introduction of the BRT}

\begin{figure}[t!]
\centering
\includegraphics[width=0.95\textwidth]{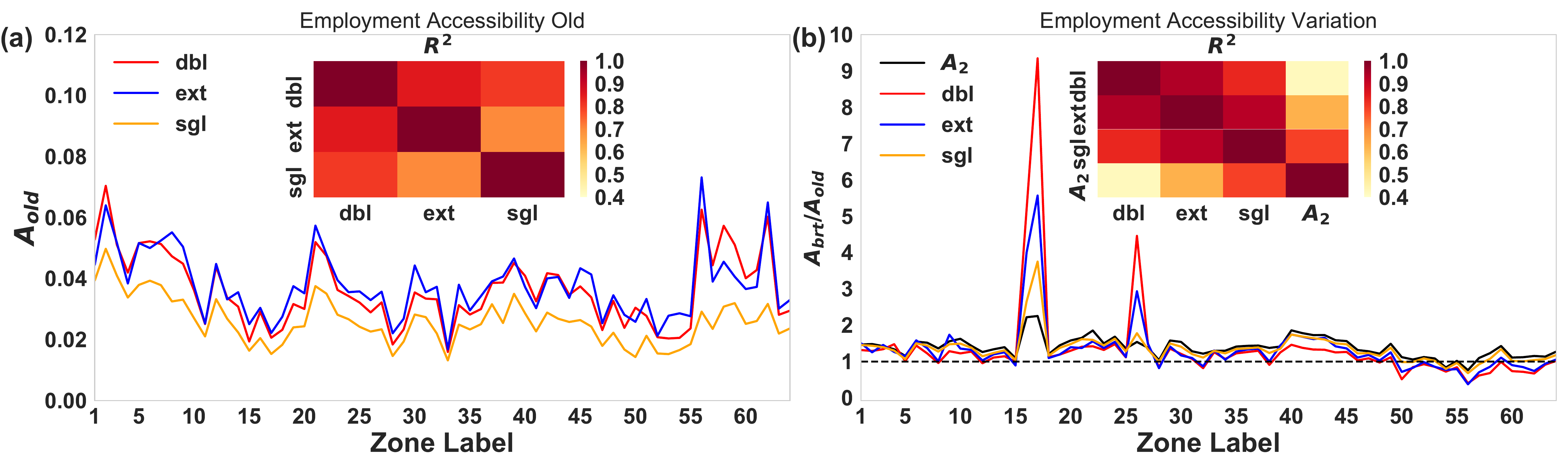}
\includegraphics[width=0.95\textwidth]{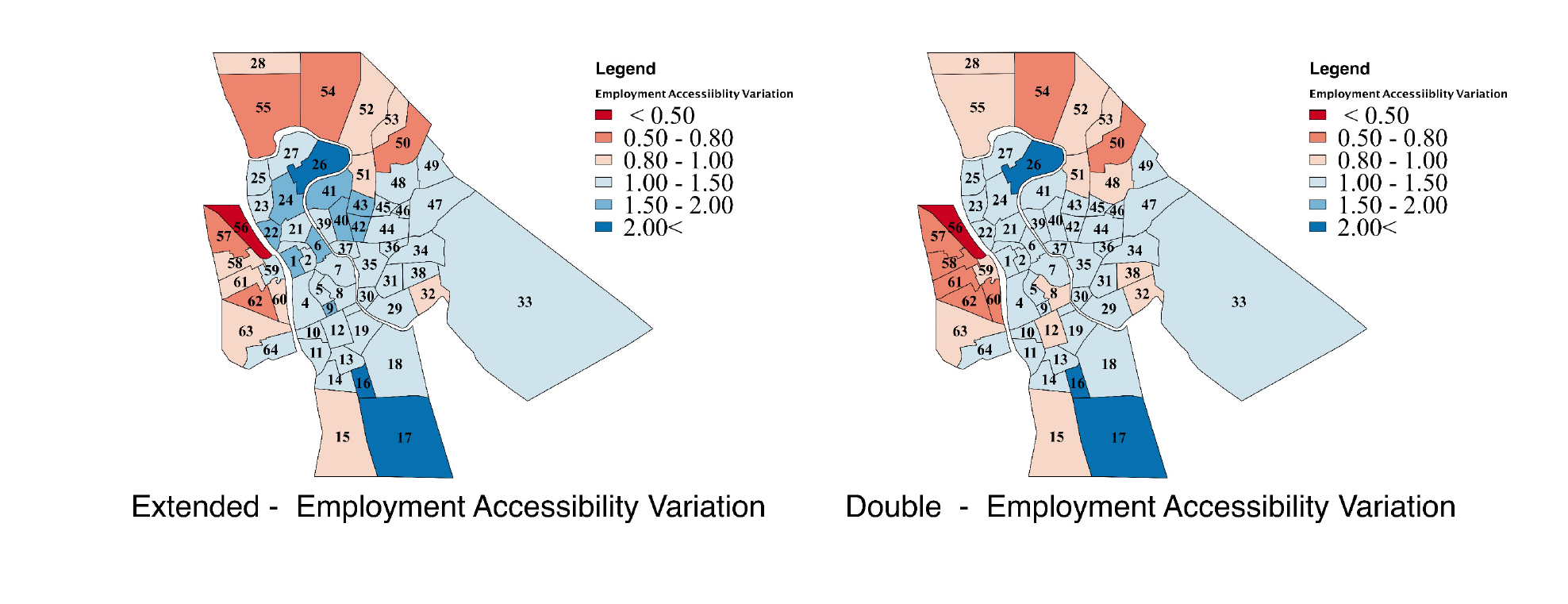}
\caption{In this figure we show the accessibility to employment with $A_\text{old}$ in \textbf{(a)} and   $\frac{A_\text{brt}}{A_\text{old}}$ in \textbf{(b)} on the top left and top right respectively, for each zone of the city of Teresina and Timon. As noted in the legends of the figures, the red and orange curves show the results obtained from the gravity model, from the doubly and singly constrained versions respectively, while the blue curve represents the results obtained with the extended radiation model. The black curve in the accessibility variation is the variation measured with respect accessibility of type 2. In the insets, we present the $R^2$ values between the various curves. The maps show the spatial distribution of $A^\text{var}$ obtained with the doubly-constrained gravity model and with the extended radiation model.}
\label{fig:3}
\end{figure}
In order to quantify the impact of the introduction of the BRT, we have measured, for all zones in Teresina,  the quantities in eq.(\ref{eq:A1}) and (\ref{eq:A2})  using the cost matrices before the intervention  $c^\text{old}_{ij}$ and after $c^\text{brt}_{ij}$. To explicitly study the variation introduced by the BRT system, we have then analysed the ratio of the two quantities
\begin{equation}
A_\text{var} = \frac{A_\text{brt}}{A_\text{old}}
\end{equation}
so that the zones with $A_\text{var}(i)>1$ are predicted to benefit from the intervention and vice-versa. We repeat this process using the gravitational and radiation approach. The results are summarized in fig(\ref{fig:3}), where we show in detail all the quantities we have discussed and the spatial distribution of the predicted impact on the city's zoning system. In the top panel we have shown $A_\text{old}$ and  $A_\text{var}$ calculated with the various models. Given the difference in the characteristic values between the accessibilities of type 1 and 2, we only compare their predictions on the $A_\text{var}$, where the variation of type 2 is represented by a black curve. We will refer to $A(T^\text{dbl})$, $A(T^\text{sng})$ and $A(T^\text{ext})$ for the accessibilities calculated using the double-constrained gravity, single-constrained gravity and the extended model respectively.  

What immediately catches the eye is the substantial agreement between the $A(T^\text{ext})$ (blue curve) and $A(T^\text{dbl})$ (red curve) both before the BRT introduction, and then in predicting the impact of its introduction. The $R^2$ values in the insets confirm this impression, with higher values between the two models than that measured between $A(T^\text{dbl})$ and $A(T^\text{sng})$. This is quite unexpected, especially considering the fact that per construction, the extended radiation model is a singly constrained model.
\begin{figure}[t!]
\centering
\includegraphics[width=\linewidth]{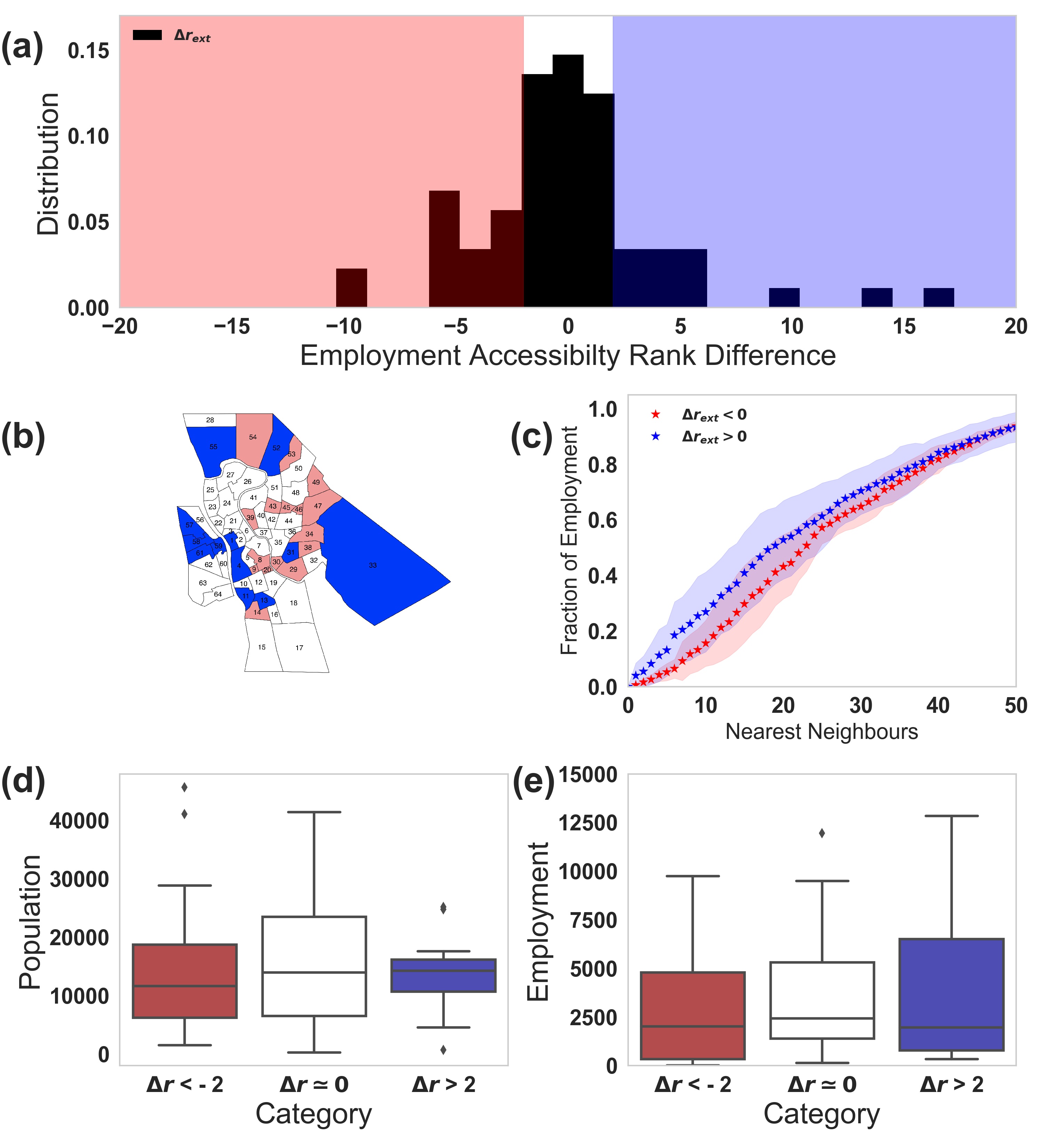}
\caption{In the top figure \textbf{(a)} we show the distribution of $\Delta r $. The blue zones highlight zones in category 1, the white area zones in category 2, and the red in category 3. In the middle figures, we show the spatial distribution of the 3 categories \textbf{(b)}, and in \textbf{(c)} for category 1 and 3 we show the average distance to employment. The shaded areas represent the standard deviation of the distribution. Finally in the bottom figures, we present for each category the distribution of population \textbf{(d)} and employment \textbf{(e)}  }
\label{fig:4}
\end{figure}
Looking at the accessibility variations $A_\text{var}$, it is clear that all three models predict two main peaks, one for zones 17 and 16, and another for zone 26. The first peak can be seen analysing both types of accessibility, $A^1$ and $A^2$ (black curve), while the peak in zone 26 is not captured by the measure of the infrastructure. The BRT map in fig.(\ref{fig:1}) shows that a new BRT stop is planned near zone 26, so a spike in its accessibility is indeed reasonable. This seems to suggest that the behavioural layer introduced through the modelling of the flows adds more information to the simple analysis of  infrastructure.  The variation of the accessibilities of type 1 predicted by the three models have a high $R^2$ value ($R^2 > 0.8$), while $A^1 (T^\text{sng})$ is the only one in good agreement with $A^2$.

To understand the spatial configuration of the results, we show a heat map of the $A_\text{var}$ on the map of the city in fig.(\ref{fig:3}), only using the doubly constrained and the extended radiation models. The maps show that the zones in the city centre are expected to benefit from the new transport network, but the main benefits are located in the south zones 16 and 17, and in the centre north zones 26 and 27. We can see from the BRT map in fig.(\ref{fig:1}) that these zones are close to where new bus stops will be positioned. The zones in the north of the city, despite being close to the new stops and served by a BRT, are expected to generate longer journeys to work, and therefore their accessibility will be lower. The zones in the district of Timon are not included in the BRT intervention, and it is therefore not surprising to find that their employment accessibilities decrease according to all the three models. 

Despite producing different aggregated travel time distributions (fig.(\ref{fig:2})), the predictions obtained with the two models are in good agreement. This is clear by looking at the high $R^2$ values obtained when comparing the curves, and its especially true for the doubly constrained and the extended radiation models. Moreover the results are reasonable  bearing in mind the spatial distribution of the BRT stops and corridors, and accessibility of type 1 seems a more appropriate measure for this study than the measure of type 2.  

\subsection{The Doubly-Constrained Gravity versus The Extended Radiation Model}

As portrayed in the introduction, one of the main objectives of this work is to analyse similarities and differences in the predictions made using the two approaches. Despite the great agreement found between the doubly-constrained and the extended radiation models, several zones show contradictory results. We will now look into the properties of these specific zones, and check if there are similarities among them. We do this by comparing the predictions on the accessibility after the BRT introduction, $A_\text{brt}(T^\text{ext})$ and $A_\text{brt}(T^\text{dbl})$, by analyzing the difference in the rank of each zone, namely
\begin{equation} 
\Delta r_i = r^\text{dbl}_i - r^\text{ext}_i
\label{eq:Dr}
\end{equation}
where $ r^\text{dbl}_i$ is the rank of zone $i$ using the gravity and  $ r^\text{ext}_i$ is the rank obtained using the radiation model. If $\Delta r_i  > 0$, this implies that $ r^\text{dbl}_i >  r^\text{ext}_i$, which means that the accessibility of zone $i$ ranks higher if we use the gravitational model to calculate the accessibilities, and vice versa. The distribution of the quantity in eq.(\ref{eq:Dr}) calculated using the two models is found in the top panel of fig.(\ref{fig:4}). The red and blue areas highlight the zones for which the predictions vary considerably ($\|\Delta r \| > 2$), which are those whose characteristics we want to analyse. We have therefore divided the zones into
\begin{itemize}
\item {Category 1: $\Delta r_i >> 0$}
\item {Category 2: $\Delta r_i \simeq 0$} 
\item {Category 3: $\Delta r_i << 0$}
\end{itemize}
Furthermore we have studied several characteristics of the zones that belong to each category: the population and employment distribution, their spatial distribution and their distance to employment. As we can see from fig.(\ref{fig:4}), the population and employment distributios are completely comparable in all three categories (d)-(e), and no precise spatial pattern emerges when projecting these categories on the city's zoning system (b). But indeed an interesting difference emerges when quantifying the distance to employment of the different categories, as we can see from  fig.(\ref{fig:4})(c). What emerges is a clear tendency of the zones of category 1 to be closer to the opportunities (blue curves) than those of category 3 (red curves). By looking at the figure, it is clear how the blue curve tends to increase faster than the red one. A zone whose closest neighbouring zones are rich in employment tends to have a larger predicted accessibility using the gravitational approach, while zones surrounded by zones with low opportunities score higher accessibility using the radiation approach. The reasons behind this result are probably to be found in the details of the equations, and as to how the methods deal with the concept of distance. We believe that this finding can be very useful in understanding future applications of the two methods.
\section{Discussion}

In this work, we have used two different approaches to test the impact of a new transportation facility. In doing so, we have achieved two results: the first of course is to enable us to cross-check our findings, but the second is comparing the gravity and radiation models models on real data and in practice. We have seen how the observed trip duration distribution can be reproduced quite well by the gravity models, in both their versions, while we also see how the radiation model fails to provide as a good and as a satisfactory description. The  extended radiation model owing to the fact that it is calibrated improves this model substantially. Interestingly the agreement between the extended radiation and the double-constrained gravity model, with different agglomerated trip duration distributions, is outstanding. This may suggest that the detailed reproduction of the observed distribution is not crucial in a city planning context with respect to measuring accessibility variations at the scale we are working at. Furthermore we have compared the estimated impact on the accessibilities, using the type 1 and 2 definitions. The type 1 accessibility, in this context, seems to be in better agreement with what one may expect. By looking at fig.(\ref{fig:3}) we see that the zones with the highest estimated improvement, zones 15, 16 and 26, have low starting accessibilities and are positioned next to a BRT stop. The modelling of flows does seem to add information that the policy maker cannot extract simply by observing the differences in the infrastructure and the trip durations. With this in mind the models that seem to better describe  mobility are the extended radiation and the doubly-constrained gravity model. We have seen how, despite considerable agreement, the predictions obtained by the two models differ for those zones with many opportunities around them and those with very few. Indeed we have seen that the zones with a neighbourhood rich in opportunities perform better with the gravity model, and vice-versa. This is an interesting result, which is potentially useful in  urban planing scenarios. We may conclude by saying that these preliminary results show that the radiation model in its extended version could be a valid alternative to study urban mobility and test new transportation networks. More research on this topic would help us better understand how the two models could support each other. 
\section{Appendix}

\subsection{Data}

The main data source is the STRANS/PMT (Teresina Transport Authority) database created through an Origin-Destination survey conducted in 2007 including the cities of Teresina and Timon \cite{Strano2008} . The database includes household conditions, personal socioeconomic information and travel diaries per person on the day before the survey was taken. The main dataset contains trip data (walking times, waiting times, travel times, mode, origin and destination zones and activities, and trip costs) combined with disaggregated demographic and socioeconomic information about the traveler (education level, income, gender, employment, age) and traveler's household data (traffic zone, comfort and deprivation variables - number of families, bedrooms, bathrooms, sewage system, access to water, energy consumption).
The survey consists of 64 traffic zones in Teresina and Timon that coincide with the cities' districts which provide the opportunity to gather sociodemographic data from the national census. In total, the dataset contains 5,177 journeys distributed across 138 households. The dataset also contains information about the number of employers, students and total population in each traffic zone. Geo-referenced data is also available for bus routes and stops for all routes in Teresina and Timon and the General Transit Feed Specification (GTFS) for the public transportation in the city. Data about Timon's buses routes were taken from Moovit App. The Bus Journeys Dataset was also taken from STRANS/PMT and contains 45,090 bus trips for a twenty-four hour interval for each bus line in 2006. The database describes single bus passengers journey through the variables: Bus Route, Bus ID, Direction, Time at Origin, Origin Bus Stop and Destination bus stop.

\subsection{Cost Function Calculations}

The effects of the new transportation network on trip distribution have been measured by calculation of the generalized costs of travel, for each pair of zones, $i\rightarrow j$. To do so we had to calculate two cost matrices, one before $\{C^\text{old}_{ij}\}$  and one after $\{C^\text{brt}_{ij}\}$ the introduction of the BRT. Each element of the matrices indicate the time necessary to travel from zone $i$ to zone $j$ using public transportation. We have collected the travel times of the old transportation network from Google Maps Directions, using the API, which provides a service to calculate directions between requested locations considering available modes of transportation.  The API request takes as inputs origins and destinations, which were taken from zone centroids coordinates, and gives an estimate of the expected time of travel. The results were stored in the Origin Destination cost matrix $C^\text{old}$. On the other hand the travel times after the BRT introduction had to be predicted. To estimate new routes, we have built a model using ArcGIS Spatial Analyst Extension for the existing bus network with the current local bus GTFS (General Transit Feed Specification, given by the Transport Authority of Teresina), and calibrated with real travel time data and the predicted travel time between pairs of zones was calculated.  Further developments for the BRT systems were introduced in the network (BRT GTFS), and the predicted travel time between pairs of zones was calculated. In this case the results were stored in the origin destination cost matrix $C^\text{brt}_{ij}$.
\newline
\newline
Data Availability: All the data used in this work can be found in this public repository https://github.com/ducciopiovani/Data-on-Teresina-, where we have stored the cost matrices and the origin and destination matrix in 3 separate files.
\newline
\newline
Competing Interests:We declare we have no competing interests. 
\newline
\newline
Authors' Contributions: All authors conceived and designed the study, analyzed and interpreted the results and drafted the manuscript. D.P. carried out the analysis, written all the codes both for the simulations and the data analysis, and coordinated the efforts. G.U. gathered and manipulated the data, built the cost matrices and wrote the first draft. 
\newline
\newline
Funding: All authors acknowledge the funding from the Engineering and Physical Sciences Research Council, grant number: EP/M023583/1.

\end{document}